\documentclass[conference]{IEEEtran}
\ifCLASSINFOpdf
\IEEEoverridecommandlockouts

\else

\fi
\usepackage{mathrsfs}

\usepackage{amsmath}
\usepackage{ntheorem}

\usepackage{makeidx}  
\usepackage{algorithm}
\usepackage{algorithmic}
\usepackage{graphicx}
\usepackage{subfigure}
\usepackage{epstopdf}
\usepackage{bm}
\usepackage{cite}
\usepackage{stfloats}
\usepackage{setspace}

\usepackage{amssymb}
\usepackage{graphicx}

\usepackage{url}

\usepackage{tabularx,booktabs}
\newcolumntype{C}{>{\centering\arraybackslash}X} 
\usepackage{lipsum}

\usepackage{makecell} 

\usepackage{graphicx}
\usepackage{color}

\usepackage{multirow}

\definecolor{dblue}{RGB}{15,89,164}

\def\BibTeX{{\rm B\kern-.05em{\sc i\kern-.025em b}\kern-.08em
		T\kern-.1667em\lower.7ex\hbox{E}\kern-.125emX}}

\usepackage[bottom=1.07 in,top=0.75in,left=0.63in,right=0.633in]{geometry} 
\usepackage{placeins} 
\usepackage{float} 

\begin{document}

\title {Near-Field  3D Localization and MIMO Channel Estimation with  Sub-Connected Planar Arrays 	\vspace{-15pt}}

\author{Kangda Zhi\textsuperscript{1}, Tianyu Yang\textsuperscript{1},  Songyan Xue\textsuperscript{2}, and  Giuseppe Caire\textsuperscript{1} \\
	
	\textsuperscript{1}Technische Universit\"{a}t Berlin, Germany, \{k.zhi, tianyu.yang,   caire\}@tu-berlin.de \\
    \textsuperscript{2}Huawei Technologies Co., Ltd., China, xuesongyan@huawei.com
	\vspace{-15pt}
}

\maketitle
%




\maketitle
\thispagestyle{empty}
\pagestyle{empty}
\begin{abstract}
	This paper investigates  the design of  channel estimation and 3D localization algorithms in a challenging scenario, where a sub-connected planar extremely large-scale multiple-input multiple-output (XL-MIMO) communicates with  multi-antenna users. In the near field, the  uplink MIMO channel is of full column rank and therefore can not be estimated effectively by applying existing codebooks that are designed for the far-field case or for the near-field case but limited to single antenna users.  To solve this problem, we propose a three-stage algorithm aided by  orthogonal matching pursuit (OMP) and sparse Bayesian learning (SBL). Specifically, we firstly partition the XL-MIMO into subarrays and use OMP to solve the compressed sensing (CS) problem about subarray channel estimation with the Discrete Fourier Transform (DFT)-based dictionary matrix. Secondly, exploiting the estimated subarray channels and employing  one-dimensional multiple signal classification (MUSIC), we estimate the central location of the user array under the Least Squares (LS) criterion. Finally, we  utilize the estimated central location to construct a refined location-aided dictionary matrix and obtain the MIMO channel estimation using SBL. Results exhibit the significant superiority of the proposed algorithm compared with several benchmarks, in terms of both the pilot overhead and estimation accuracy.
\end{abstract}
%
%
%

\IEEEpeerreviewmaketitle

\section{Introduction}
To fulfill the stringent data rate requirement for the upcoming sixth generation (6G) network, a key approach consists of increasing the number of antennas, leading to the emerging technology called extremely large-scale multiple-input multiple-output (XL-MIMO)\cite{liu2025evaluating,Lu2024Tutorial}.  Recently, XL-MIMO has been explored through a broad body of research activity and experimental testing, with emphasis on the aspects of channel modeling and estimation\cite{han2020channel,cui2022channel,zhou2025mixed,zhi2025near}, codebook design\cite{wu2022multiple}, localization\cite{he20233d,pan2023ris22}, channel measurement\cite{Tang2024XLMIMO},  low-complexity design\cite{zhi2024performance}, and so on.

To fully leverage the promising benefits brought by XL-MIMO, effective acquisition of the channel state information is essential. Unlike the planar-wave-based far-field channel where the array steering vectors depend only on angles of arrival/departure, the spherical-wave-based near-field channel involves array steering vectors that depend both on angles and distance. This additional distance-domain information poses new challenges to the design of beam codebooks and channel estimation algorithms. 

When the user device is equipped with a single antenna, polar-domain and spherical-domain codebook were proposed in \cite{cui2022channel,wu2022multiple} for estimating the single-input multiple-output (SIMO) channel  with uniform linear array (ULA) and uniform planar array (UPA) at the base station (BS),  respectively, by sampling the distance domain non-uniformly. When the user device has multiple antennas, the problem becomes more challenging since the near-field LoS MIMO channel is of full rank and can not be decomposed  as the product of two rank-one channels. The estimation of near-field LoS MIMO channel has been treated only by a few works. Specifically, the authors in \cite{lu2023near} proposed to estimate the line-of-sight (LoS) MIMO channel between two ULAs with on-grid search and gradient descent. The authors in \cite{liu2025sensing} considered searching the user location using a holography-inspired reconstruction method and estimated the MIMO channel between ULAs with a location-based eigen-dictionary. By introducing a  Vandermonde windowing matrix and exploring the structure of antenna-pair distances, the MIMO channel between ULAs was estimated in \cite{shi2024double}. On the other hand, for localization problem, a highly influential work \cite{hua20243d} investigated a 3D multi-target localization system leveraging multiple IRSs and introduced a pioneering IRS-adaptive sensing protocol. It developed a novel multi-beam IRS scheme to effectively overcome the failure of target localization in scenarios where the BS-IRS link is dominated by LoS propagation.

However, all the above-mentioned work considered estimating high-rank MIMO channels between ULAs, which cannot be directly applied to the more general scenario when the BS is equipped with a 2D UPA. The estimation of LoS MIMO channel between multi-antenna users and UPA BS is more challenging since it introduces one more parameter, i.e., the elevation angles, which aggravates the required size of codebook and complexity of estimation algorithms, especially considering that the existing proposals for the ULA case are  already complex enough. Besides, the coupling between horizontal and vertical antennas on UPA will make some key approximations dedicated  for ULAs invalid, which further complicates the problem. Furthermore, most of the existing work focused on the fully digital array or fully-connected hybrid array, while effective design with respect to the low-cost sub-connected hybrid array has not gained enough attention yet. 

Against the above background,  this work proposes a novel three-stage joint 3D localization and MIMO channel estimation method in the challenging scenario where multi-antenna users communicate with a planar sub-connected XL-MIMO. We first estimate subarray channels with simple Discrete Fourier Transform (DFT) dictionaries in stage 1. Then, we combine subarray channels to localize the central position of the user array in stage 2   under the Least Squares (LS) criterion.  In stage 3, we construct a small-size location-aided dictionary and estimate the MIMO channel. The proposed method requires short pilot sequences and small-size dictionaries while achieving superior performance compared with several benchmarks.

\section{System Model}

\begin{figure}
	\setlength{\abovecaptionskip}{-10pt}
	\setlength{\belowcaptionskip}{-10pt}
	\centering
	\includegraphics[width= 0.45\textwidth]{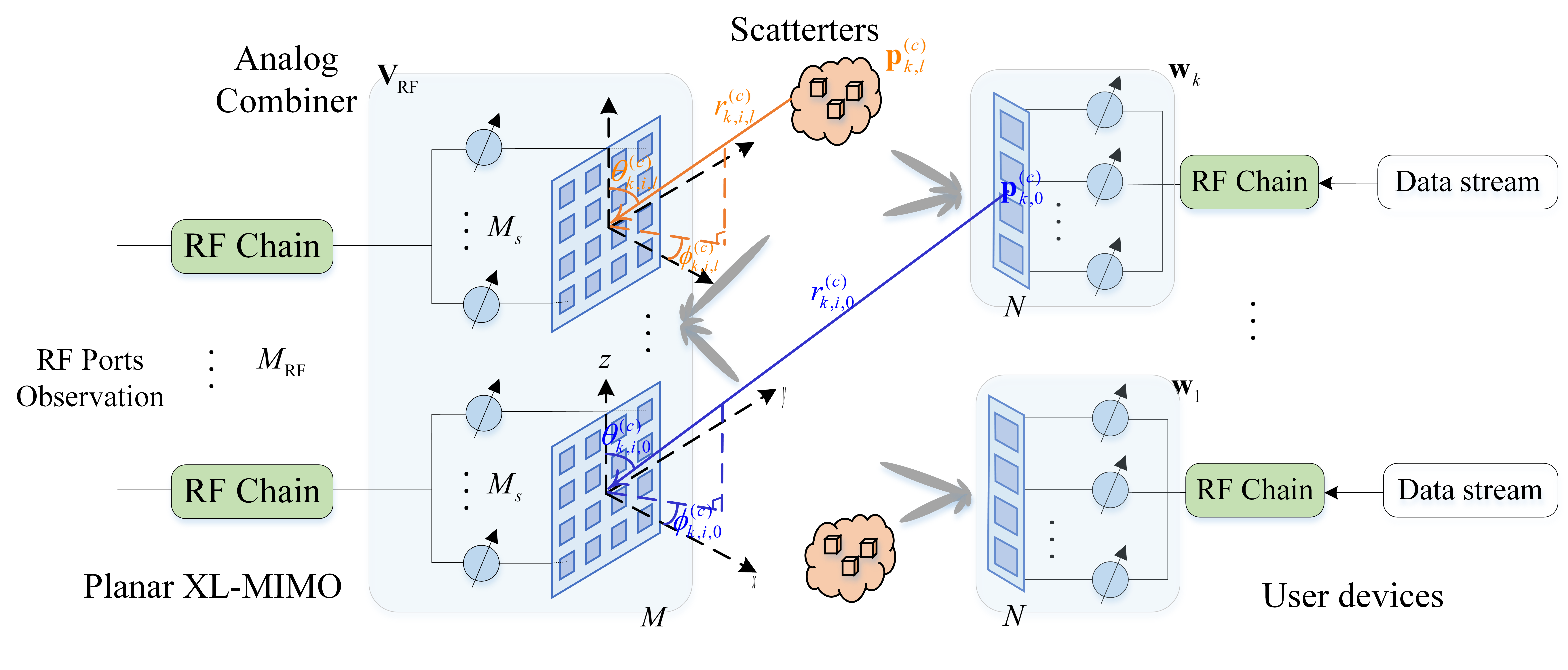}
	\caption{Illustration of the considered sub-connected MIMO system.}
	\label{figure1}
	\vspace{-15pt}
\end{figure}

As shown in Fig. \ref{figure1}, an XL-MIMO BS  is considered to communicate with $K$ multi-antenna users. To reduce hardware complexity, a sub-connected structure is employed at both the BS and the user. On the BS array,  $M$ antenna elements are connected to $M_{{\mathrm {RF}}}$ radio-frequency (RF) chains, where $M_{{\mathrm {RF}}}\ll  M$ and $M_s=M/M_{{\mathrm {RF}}}$. Given the limited device space,  each user is equipped with  a ULA of $N\ll M$ antennas connected by a single RF chain.    The BS array is an UPA such that the physical form factor of the XL-MIMO array is still manageable and 3D localization is enabled.  For UPA at the BS, it assumed that $M=M_h M_v$, where $M_h$ ($M_v$) and $\Delta_h$ ($\Delta_v$)  represent the number of antennas and antenna spacing at horizontal  (vertical) directions, respectively.

\subsection{Channel Model}

The propagation environment between user $k$  and the BS is modeled as a parametric channel with $L_k$ paths. Given the large number of antennas at the BS, the  user and scatterings are likely located in the near field of the BS. Therefore, near-field channel model should be considered. The uplink channel $\mathbf{H}_k\in \mathbb{C}^{M \times N}$ from user $k$ to the BS    with $L_k+1$ paths can be expressed as follows 
\begin{align}\label{channel_H1}
&	\mathbf{H}_k  =\mathbf{H}_k ^{\rm los} + \mathbf{H}_k ^{\rm nlos}  \nonumber\\
	&=\mathbf{H}_k ^{\rm los}   + \sum\nolimits_{l=1}^{L_k}  \alpha_{k,l}
	{\mathbf{h}}_{k,l}    (r_{k,l}, \theta_{k,l} , \phi_{k,l} ) 
	({\mathbf{h}}_{k,l} ^ {\mathrm{f}}     (\psi_{k,l})      )^H 
\end{align}
where $  \mathbf{H}_k ^{\rm los} (m,n)  = \frac{1}{r_{k,m,n}}e^{-j\frac{2\pi}{\lambda} r_{k,m,n}} $  denotes the full-rank near-field LoS channel between the $n$-th antenna of user $k$  and the $m$-th antenna of the BS, specified by antenna-pair distance  $  r_{k,m,n}  $.
The paths with $l\in [1,L_k]$ correspond to the  non-LoS (NLoS) paths from scatters.  $\alpha_{k,l} $ includes the distance-dependent pathloss and small-scale fading coefficient corresponding to the $l$-th path for user $k$.
Parameters  $r_{k,l}$ , $\theta_{k,l} $, and $\phi_{k,l} $, which constitute the near-field steering vector $ 	{\mathbf{h}}_{k,l}  $, denote the distance, elevation angle of arrival (AoA), and azimuth AoA, from the $l$-th scatter to the array center of the BS, respectively. $\psi_{k,l}$ denote the far-field elevation angle of departure (AoD) from user $k$ to the $l$-th scatter due to the limited number of antennas $N$.

\subsection{Received Signal Model}
The received signal is formulated in a general case to be compatible with conventional channel estimation methods.  Assume that the entire uplink pilot transmission period occupies  $BT$ time slots, which are organized into $B$ blocks, each containing $T$ time slots.  At every block $\tau  \in[1:B]$,  each user $k$ repeats transmitting pilot $\mathbf{w}_{k,\tau} s_{k} $ for $T$ times, where $s_{k}  = \sqrt{p_k}$ is the transmitted signal with power $p_k$ and $\mathbf{w}_{k,\tau} \in\mathbb{C}^{N\times 1}$ is the analog precoder  with constant  modulus constraints $\left| \left[ \mathbf{w}_{k,\tau} \right]_n\right|\!\!=\!\! \frac{1}{\sqrt{N}}$, $1\le n \le N$. Meanwhile, in every block $\tau$, the BS tries $T$ different receiving patterns $ \mathbf{V}_{{\mathrm {RF}},t} \in\mathbb{C}^{M_{\mathrm {RF}}\times M}$, $t\in [1:T]$, which can be expressed as\cite{zhi2025holographic}
\begin{align}
	\mathbf{V}_{{\mathrm {RF}},t} = \mathrm{blkdiag} \left\{   \mathbf{v}_{{\mathrm {RF}},1, t}^{\rm T },   \dots, \mathbf{v}_{{\mathrm {RF}}, M_{\mathrm{RF}},t}^{\rm T },       \right\},
\end{align}
in which $\mathbf{v}_{{\mathrm {RF}}, m_{\mathrm{rf}},t}   \in\mathbb{C}^{M_s\times 1}  $ is the analog combining vector of phase shifter network connected to the $m_{\rm rf}$-th RF chain, with constant modulus $\left|\left[ \mathbf{v}_{{\mathrm {RF}}, m_{\mathrm{rf}},t} \right]_{m_s}\right|=\frac{1}{\sqrt{M_s}}$.  


 In time $t$ of block $\tau$, the signal received by the BS\footnote{We assume mutual-orthogonal pilot sequences are transmitted by $K$ users to the BS, e.g., using orthogonal time or frequency resources. Therefore, channel estimation  is independent among users and without loss of generality, an arbitrary user $k$ will be considered. } is given by $\mathbf{y}_{k,\tau,t} =   \mathbf{V}_{{\mathrm {RF},t}} ( \sqrt{  p_k}  \mathbf{H}_k \mathbf{w}_{k,\tau}   +  \mathbf{n}_{\tau,t})$.  Then, after $T$ successive combining beams in block $\tau$, the $T M_{\rm RF} \times  1$ received signal at the BS RF ports can be expressed as
\begin{align}\label{received_signal}
\mathbf{y}_{k,\tau} \!\!=\! [\mathbf{y}_{k,\tau,1}^{\rm T} ,\ldots,\mathbf{y}_{k,\tau,T}^{\rm T}  ]^{\rm T} \!=\!\!  \sqrt{  p_k} \;\mathbf{V}_{{\mathrm {RF}}} \mathbf{H}_k \mathbf{w}_{k,\tau}   \!\!+\!  \widehat{\mathbf{V}}_{{\rm RF}}   \mathbf{n}_{\tau},\!
\end{align} 
where $\mathbf{V}_{{\mathrm {RF}}} = [\mathbf{V}_{{\rm RF},1}^{\rm T}, \dots,  \mathbf{V}_{{\rm RF}, T}^{\rm T}]^{\rm T} \in\mathbb{C}^{T M_{\rm RF} \times  M}$ and $\widehat{\mathbf{V}}_{{\rm RF}} = \mathrm{blkdiag}\left\{ \mathbf{V}_{{\rm RF},1} , \dots,  \mathbf{V}_{{\rm RF}, T}   \right\}  \in\mathbb{C}^{T M_{\rm RF} \times  TM}$ are the aggregated analog combiners and $\mathbf{n}_{\tau}\sim \mathcal{C N}\left(   \mathbf{0}, \sigma^2 \mathbf{I}_{TM}  \right) \in\mathbb{C}^{T M  \times  1}$ represents the Gaussian noise vector at the BS antenna elements.
By concatenating   transmitting beam patterns after  $B$   blocks, the  $T M_{\mathrm{RF}} \times B $  received signal  can be constructed as
\begin{align}\label{Y}
\mathbf{Y}_k=  \sqrt{  p_k} \;\mathbf{V}_{{\rm RF}} \mathbf{H}_k \mathbf{W}_k   +  \widehat{\mathbf{V}}_{{\rm RF}} \mathbf{N},
\end{align}
where $\mathbf{Y}_k =[ \mathbf{y}_{k,1} , \dots, \mathbf{y}_{k,B}   ]$, $\mathbf{W}_k =[\mathbf{w}_{k,1}, \dots, \mathbf{w}_{k,B} ] \in\mathbb{C}^{N\times B}$, and $\mathbf{N} = [\mathbf{n}_1, \dots, \mathbf{n}_B ] \in\mathbb{C}^{TM\times B}$.

The estimation of $\mathbf{H}_k$ from (\ref{Y}) is very challenging due to the following reasons: 1) the LoS MIMO channel $\mathbf{H}_k^{\rm los}$ is of full column rank and can not be decomposed into the sum of product of some rank-1 channels as simple as $\mathbf{H}_k^{\rm nlos}$, making the design of dictionary matrix difficult; 2) the structure of planar array and near-field feature lead to a large dimension of unknown parameters, i.e.,  $ r_{k,m,n}$ for LoS and $ r_{k,l}, \theta_{k,l}, \phi_{k,l}  $ for NLoS, making the size of potential dictionary matrix or search space of multiple signal classification (MUSIC)  methods prohibitive; 3) the sub-connect hybrid architecture of the arrays bring challenges to the design of an effective sensing matrix in the compressed sensing (CS) problem.

\section{ Antenna-Wise Estimation  }\label{section_ce}
The feasibility of conventional methods is mainly limited by the full-rank feature of  LoS MIMO channel $\mathbf{H}_k^{\rm los}$. To overcome this difficulty, a possible way is to estimate the SIMO channel from each separated antenna of the user to the BS array, i.e., a column of $\mathbf{H}_k $, instead of simultaneously estimating the whole MIMO channel. 

Letting $B=N$,   we can design the aggregated analog precoder  $\mathbf{W}_k$  as the DFT matrix $ \boldsymbol{\mathcal{F}}_{N} $, i.e., $\left[\mathbf{W}_k\right]_{n,b} = \frac{1}{\sqrt{N}}\exp\{  -j2\pi  \frac{(n-1)(b-1)}{N}   \}$, leading to  the property of $ \mathbf{W}_k\mathbf{W}_k^H = \mathbf{I}_N$.   Denote   $\mathbf{H}_k  = [\mathbf{h}_k^{(1)}, \cdots, \mathbf{h}_k^{(N)}]$ where  $ \mathbf{h}_k^{(n)}  \in \mathbb{C} ^{M \times 1}$ represents the channel from the $n$-th, $1\leq n \leq N$, antenna of the user to the BS array. Then, by multiplying receiving signal $\mathbf{Y}_k$ in (\ref{Y}) with $\mathbf{W}_k^H$ and extracting data from the $n$-th column,  we can construct the observation vector with respect to the SIMO channel  $  \mathbf{h}_k^{(n)}   $ as follows
\begin{align}\label{ob_conv}
\mathbf{y}_k^{n} =\left[ \mathbf{Y}_k\mathbf{W}_k^H \right]_{:,n} 
=  \sqrt{  p_k} \;\mathbf{V}_{{\rm RF}} \mathbf{h}_k^{(n)}    + {\mathbf{n}}_{n} ,\;\; 1\le n \le N,
\end{align}
where $ {\mathbf{n}_n} = \widehat{\mathbf{V}}_{{\rm RF}} \mathbf{N}  \left[ \boldsymbol{\mathcal{F}}_N\right]_{n,:}^H$ and  $\mathbb{E}\left\{   {\mathbf{n}_n} {\mathbf{n}_n}^H     \right\} = \sigma^2    \widehat{\mathbf{V}}_{{\rm RF}} \widehat{\mathbf{V}}_{{\rm RF}}^H$. 

It can be found that the  SIMO channel $ \mathbf{h}_k^{(n)}   $ can be parameterized by the following tractable form:
\begin{align}\label{channel_Hn}
	\mathbf{h}_k^{(n)}   &\!=  \! {\mathbf{h}}_{k,0} (r_{k,0}^{(n)}, \theta_{k,0}^{(n)}, \phi_{k,0}^{(n)})  
	\!+\! \sum\nolimits_{l=1}^{L_k} \alpha_{k,l}^{(n)}
	{\mathbf{h}}_{k,l}     (r_{k,l}^{(n)}, \theta_{k,l}^{(n)}, \phi_{k,l}^{(n)}) \nonumber \\
	& \approx \sum\nolimits_{l=0}^{L_k} \alpha_{k,l}^{(n)}
	{\mathbf{h}}_{k,l}     (r_{k,l}^{(n)}, \theta_{k,l}^{(n)}, \phi_{k,l}^{(n)}),
\end{align}
where $ r_{k,l}^{(n)}$, $\theta_{k,l}^{(n)}$, $\phi_{k,l}^{(n)} $ denote the distance,  elevation AoA, and azimuth AoA from the $n$-th antenna of the user ($l=0$) or  from scatterer $l >0$ that reflects signal from antenna $n$ to the center of the planar BS array, respectively. Besides, we have 
\begin{align}\label{hkl}
	[	{\mathbf{h}}_{k,l}     (r_{k,l}^{(n)}, \theta_{k,l}^{(n)}, \phi_{k,l}^{(n)}) ]_m = \mathrm{exp}\big(  - j \frac{2\pi}{\lambda}   
	\left\|         \mathbf{p}_{\mathrm{bs}}^{(m)} -\mathbf{p}_{k,l}^{(n)}   \right\|  
	\big)
\end{align}
where $   \mathbf{p}_{\mathrm{bs}}^{(m)} $ and $ \mathbf{p}_{k,l}^{(n)} $ denote the location of the $m$-th antenna of the BS and the $n$-th antennas of the  scatterers/user, respectively.

Exploiting  the form (\ref{channel_Hn}), we can  represent the near-field SIMO channel using the 3D spherical-domain dictionary designed for the planar array\cite{wu2022multiple}, and apply conventional CS methods to estimate the $N$ SIMO channels one by one, based on the observation $ \mathbf{y}_k^{n} $ in (\ref{ob_conv}), after $TN$ time slots ($B=N$).

\section{Three-Stage Algorithm}
Conventional methods proposed in Section \ref{section_ce} suffer from two drawbacks. Firstly, it requires a long period $TB $ to transmit the pilot. Secondly, due to the large number of antennas, the size of  spherical dictionary matrix would be prohibitively huge to  sample  the 3D space with fine grids, which is not favorable for computation efficiency and algorithm convergence.

To tackle these challenges, we  propose a three-stage algorithm that requires a small-size dictionary matrix and a short period of pilot transmission, by leveraging the capability of 3D localization for planar XL-MIMO and the potential of location information in channel estimation. The idea is to first partition the XL-MIMO array into small subarrays and  estimate the  SIMO channel from the center of user array to each subarray using DFT dictionaries.  Then,  the subarray channel information is combined to localize the central position of the user. Finally, we construct a small-size location-aided dictionary matrix and estimate the MIMO channel. By exploring the problem property,   the proposed algorithm can achieve superior estimation accuracy within one block ($B=1$).

\subsection{First Stage: Subarray-Wise Channel Estimation}
We divide the whole XL-MIMO  into $I$ tiles and re-organize the subarray-wise received signal in $B=1$ block. Design the transmit beam from the user  as $\mathbf{w}_{k,1} = \mathbf{1}_N /\sqrt{N} $. Denote   the number of antennas and  RF chains of the $i$-th subarray as $M_i = M_{i,\mathrm{h}}  M_{i,   \mathrm{v}  } $  and $M_{\mathrm{rf},i}$, respectively. Then, the  signal $ \mathbf{y}_{k,i,t} \in \mathbb{C} ^{M_{\mathrm{rf},i} \times 1}  $ received by the $i$-th subarray at the $t$-th time slot can be processed as
\begin{align}
		&\mathbf{y}_{k,i,t} =  \sqrt{ p_k/N} \;\mathbf{V}_{{\rm RF},i,t} \mathbf{H}_{k,i}   \mathbf{1}_N+  \mathbf{V}_{{\rm RF},i,t} \mathbf{n}_{i,t}\\
		& =  \sqrt{ p_k/N} \;\mathbf{V}_{{\rm RF},i,t}  \sum\nolimits_{n=1}^{N} \mathbf{h}_{k,i}^{(n)}   +  \mathbf{V}_{{\rm RF},i,t} \mathbf{n}_{i,t} \nonumber  \\
		& \overset{(a)}{\approx} \sqrt{ p_k/N} \;\mathbf{V}_{{\rm RF},i,t} \;  \bar{\alpha}_{k,i}^{(c)} \mathbf{h}_{k,i}^{(c)}  +  \mathbf{V}_{{\rm RF},i,t} \mathbf{n}_{i,t}  \nonumber\\
		& =  \! \! \sqrt{ \frac{p_k}{N}}  \mathbf{V}_{{\rm RF},i,t}    \!  \sum_{l=0}^{L_{k,i}}   \bar{\alpha}_{k,i,l} {\mathbf{h}}_{k,i,l}   (r_{k,i,l}^{(c)}, \theta_{k,i,l}^{(c)}, \phi_{k,i,l}^{(c)}) 
		\! +\! \!   \mathbf{V}_{{\rm RF},i,t} \mathbf{n}_{i,t}, \nonumber
\end{align}
where (a) approximates the sum of  SIMO channel as a single effective SIMO channel from the center of user array to the BS, by exploiting the feature that the size of array within the user device is  tiny. $r_{k,l}^{(c)}$, $\theta_{k,l}^{(c)}$, and $\phi_{k,l}^{(c)}$ denote the distance and AoAs from the center of user/scatters to the BS, respectively.  $  \bar{\alpha}_{k,i,l} $ represents the effective aggregated weight for path $l$. $ \mathbf{V}_{{\rm RF},i,t}= \mathrm{blkdiag}  \{     \mathbf{v}_{1,t}^{\rm T },  \cdots ,  \mathbf{v}_{   M_{\mathrm{rf},i}    ,    t}^{\rm T }     \}  \in  \mathbb{C}^{ M_{\mathrm{rf},i} \times M_{ i} } $ denotes the sub-connnected RF precoding matrix of the $i$-th subarray.

By collecting the received pilot signal across a total of $T$ time slots, the overall observation vector at subarray $i$ can be obtained as $ \mathbf{y}_{k,i} = [   \mathbf{y}_{k,i,1}^{\rm T }, \cdots, \mathbf{y}_{k,i,T}^{\rm T }    ]^{\rm T }   \in  \mathbb{C}^{ TM_{\mathrm{rf},i} \times 1} $ where
\begin{align}\label{observation_i}
	\mathbf{y}_{k,i} \!=  \!\sqrt{ \frac{p_k}{N}} \mathbf{V}_{{\rm RF},i}   \sum_{l=0}^{L_{k,i}}  \bar{\alpha}_{k,i,l} {\mathbf{h}}_{k,i,l}   (r_{k,i,l}^{(c)}, \theta_{k,i,l}^{(c)}, \phi_{k,i,l}^{(c)}) \!+\!    \mathbf{n}_{i} ,
\end{align}
and  
\begin{align}\label{V_RF_i}
	\mathbf{V}_{{\rm RF},i}  &= [   \mathbf{V}_{{\rm RF},i,1} ^{\rm T } , \cdots, \mathbf{V}_{{\rm RF},i, T} ^{\rm T }    ]^{\rm T }  \in  \mathbb{C}^{ TM_{\mathrm{rf},i} \times M_{ i} },\\
		\mathbf{n}_{i}   &= [  \mathbf{n}_{i,1}  ^{\rm T }  \mathbf{V}_{{\rm RF},i,1}^{\rm T }  , \cdots,  \mathbf{n}_{i,T} ^{\rm T } \mathbf{V}_{{\rm RF},i,T}^{\rm T }     ]^{\rm T } .
\end{align}
 
We propose to use a CS method to estimate $ {\mathbf{h}}_{k,i,l}  $ from $ \mathbf{y}_{k,i}  $. Thus, we construct a tractable dictionary matrix in the following. Let the BS array be located on the $ yoz $ plane and let its center be the origin of the coordinate system.  As in (\ref{hkl}), denoting the center of the $i$-th subarray on the BS and scatters/user as $\mathbf{p}_{\mathrm{bs},i}$ and $\mathbf{p}_{k,l}^{(c)}$, respectively,  we have $ 	\mathbf{p}_{k,l}^{(c)}  =  \mathbf{p}_{\mathrm{bs},i}  +   r_{k,i,l}^{(c)} \mathbf{k}_{k,i,l}^{(c)} $ where
 \begin{align}
 \!	\mathbf{k}_{k,i,l} ^{(c)} \!=\! [\sin \theta_{k,i, l}^{(c)}  \cos \phi_{k,i,l}^{(c)} ,   \sin \theta_{k,i, l}^{(c)}  \sin \phi_{k,i, l}^{(c)} ,      \cos \theta_{k,i, l}^{(c)}  ]^{\rm T } \!.  
 \end{align}
 
 Then, for subarrays with small apertures, the distance between $  \mathbf{p}_{k,l}^{(c)} $ and the $(m_{i,h},m_{i,v})$-th antenna of  subarray $i$, 
  where $ - {(M_{i,\mathrm{h}}-1)}/{2} \le m_{i,h} \le {(M_{i,\mathrm{h}}-1)}{2} $ and $ - {(M_{i,\mathrm{v}}-1)}/{2} \le m_{i,v} \le   {(M_{i,\mathrm{v}}-1)}/{2} $,  can be approximated as \cite{zhi2024performance}
 \begin{align}
 		& \left\|   \mathbf{p}_{k,l}^{(c)}  -  \left(  \mathbf{p}_{\mathrm{bs},i}  + [0, m_{i,h}\Delta_h  ,   m_{i,v} \Delta_v ] \right)     \right\|_2         \\
 		&\approx  r_{k,i,l}^{(c)} - (   \sin \theta_{k,i, l}^{(c)}  \sin \phi_{k,i, l}^{(c)}     m_{i,h} \Delta_h   +  \cos \theta_{k,i, l}^{(c)}     m_{i,v}\Delta_v  ). \nonumber
 \end{align}
 Thus,  with a scalar $\gamma_{k,i,l}=\exp\{ - j \frac{2\pi}{\lambda} r_{k,i,l}^{(c)}    \}$,  the spherical steering vector $ 	{\mathbf{h}}_{k,i,l}   $ of subarray $i$  can be approximated by
 \begin{align}\label{gsdfgsd}
 	\begin{aligned}
 		&	{\mathbf{h}}_{k,i,l}      (r_{k,i,l}^{(c)}, \theta_{k,i,l}^{(c)}, \phi_{k,i,l}^{(c)})  \approx   \gamma_{k,i,l}	{\mathbf{h}}_{k,i,l}      ( \theta_{k,i,l}^{(c)}, \phi_{k,i,l}^{(c)})  \\
 		& = \gamma_{k,i,l} \mathbf{a}_{M_{i,\mathrm{h}}}^{}  \left(      \sin  \phi_{k,i,l}^{(c)}         \sin \theta_{k,i,l}^{(c)}    \right)  
 		\otimes
 		\mathbf{a}_{M_{i,   \mathrm{v}  }}^{}  \left(      \cos   \theta_{k,i,l}^{(c)}   \right),
 	\end{aligned}
 \end{align}
 where
 \begin{align}
 \left[	 \! \mathbf{a}_{M_{i,\mathrm{e}}}^{}   \! \left(     \varpi    \right)  \right]_m   \!\! =  \!  \exp   \!    \!   \left\{ j\frac{2\pi }{\lambda} \varpi  \!   \left(       \! -\frac{M_{i,\mathrm{e}}-1}{2}   \! + m   \! -1   \!    \right) \Delta_{\rm e}   \!   \right\}.
 \end{align}
 To represent channels in (\ref{gsdfgsd}), we can sample the horizontal and elevation angular space by $Z$  grid points, respectively, where
 \begin{align}
 	\sin  \phi_{z_1}         \sin \theta_{z_1}  &= ({2{z_1}-Z-1})/{Z} , \; \; {z_1} = 1, \cdots, Z,\\
 	\cos   \theta_{z_2} &=  ({2{z_2}-Z-1})/{Z},  \;\;     {z_2} = 1, \cdots, Z,
 \end{align}
 and construct the dictionary matrix $  \mathbf{A}   \in   \mathbb{C}^{ M_{i} \times Z^2}$  as
 \begin{align}
 	\mathbf{A} \!=\! \{
 	\mathbf{a}_{M_{i,\mathrm{h}}}^{} \!\! \left(    \sin  \phi_{z_1}         \sin \theta_{z_1}     \right)  
 	\! \otimes \!
 	\mathbf{a}_{M_{i,   \mathrm{v}  }}^{}  \! \!\left(     \cos   \theta_{z_2}   \right): \!  1\le z_1, z_2 \le Z  \}  .
 \end{align}
 Then, we have 
 \begin{align}\label{Ax}
 	 \sum\nolimits_{l=0}^{L_{k,i}}  \bar{\alpha}_{k,i,l} {\mathbf{h}}_{k,i,l}   (r_{k,i,l}^{(c)}, \theta_{k,i,l}^{(c)}, \phi_{k,i,l}^{(c)})
 	\approx
 	\mathbf{A} \mathbf{x}_{k,i} ,
 \end{align}
 where $  \mathbf{x}_{k,i}    \in   \mathbb{C}^{  Z^2 \times 1}$ is an approximately $(L_{k,i}+1)$-sparse vector including the effective gains of all paths. Substituting (\ref{Ax}) into (\ref{observation_i}), the equivalent observation vector for subarray $i$ can be expressed as
 \begin{align}\label{y_k_i}
 	\mathbf{y}_{k,i}  \approx  \sqrt{ p_k/N} \;\mathbf{V}_{{\rm RF},i} \mathbf{A} \mathbf{x}_{k,i}   +    \mathbf{n}_{i} =   \bar{\mathbf{A}}_i \mathbf{x}_{k,i}   +    \mathbf{n}_{i}
 \end{align}
 where $\bar{\mathbf{A}}_i  \triangleq    \sqrt{ p_k/N}\mathbf{V}_{{\rm RF},i} \mathbf{A}  $ is the equivalent sensing matrix.

Therefore, we can formulate the following CS problem
\begin{align}
	\begin{aligned}
	\hat{\mathbf{x}}_{k,i}  = 	  \arg\min_{   \mathbf{x}_{k,i}   }  \left\|\mathbf{x}_{k,i}\right\|_0
		\text { s.t. }   {\mathbf{y}}_{{k,i}}  =
		\bar{\mathbf{A}}_i \mathbf{x}_{k,i},
	\end{aligned}
\end{align}
which can be solved, for example, by OMP, with the complexity mainly of $\mathcal{O}\left( I T M_{\mathrm{rf},i}  Z^2 \right)$.

\subsection{Second Stage:  Localization}
In this stage, we extract the LoS signal AoAs about each subarray and localize the user. Since  subarrays are closely co-located, subarray channels estimated by low-complexity OMP methods    inevitability exist grid-related  quantization errors and  will blur the localization. Thus, we first apply the low-complexity 1D MUSIC algorithm to refine the azimuth and elevation AoAs at each subarray based on the estimated subarray channel $ \sum_l \hat{\bar{\alpha}}_{k,i,l}    \widehat {\mathbf{h}}_{k,i,l} ^{(c)}     ( \hat{\theta}_{k,i,l}^{(c)}, \hat{\phi}_{k,i,l}^{(c)}) =\mathbf{A} \hat{\mathbf{x}}_{k,i}  $. 

Specifically, we calculate the covariance matrix of $i$-th subarray channel by 
\begin{align}
\mathbf{C}_i  \!=\!\! \sum_{l_1, l_2}\hat{\bar{\alpha}}_{k,i,l_1}   \widehat {\mathbf{h}}_{k,i,l_1} ^{(c)}  \!   \left\{ \hat{\bar{\alpha}}_{k,i,l_2}   \widehat {\mathbf{h}}_{k,i,l_2} ^{(c)} \right\}  ^H \!\!\! \approx  \left|\hat{\bar{\alpha}}_{k,i,0} \right|^2 \mathbf{C}_{i,h} \! \otimes \! \mathbf{C}_{i,v}
\end{align}
where
\begin{align}
 \mathbf{C}_{i,h } & =  \mathbf{a}_{M_{i,\mathrm{h}}}^{}  \left(   \left[  \hat{ \mathbf{k}}_{k,i,0} ^{(c)} \right]_2 \right)   \mathbf{a}_{M_{i,\mathrm{h}}}^{H}  \left(   \left[  \hat{ \mathbf{k}}_{k,i,0} ^{(c)} \right]_2 \right),   \\
  \mathbf{C}_{i,v } & =
\mathbf{a}_{M_{i,   \mathrm{v}  }}^{}  \left(      \left[  \hat{ \mathbf{k}}_{k,i,0} ^{(c)} \right]_3       \right)      \mathbf{a}_{M_{i,   \mathrm{v}  }}^{H}  \left(      \left[  \hat{ \mathbf{k}}_{k,i,0} ^{(c)} \right]_3       \right).
\end{align}
We can extract $ \mathbf{C}_{i,h } $ and $\mathbf{C}_{i,v }$ from $ \mathbf{C}_i  $ by exploiting the property that the diagonal elements of $ \mathbf{C}_{i,h } $ and $\mathbf{C}_{i,v }$ are all ones. This means that $\mathbf{C}_{i,v }$ makes up the diagonal block of $  \mathbf{C}_i  $ while the elements of $ \mathbf{C}_{i,h } $  are distributed at the diagonal elements of each $M_{i,   \mathrm{v}  }\times M_{i,   \mathrm{v}  }$ block of $ \mathbf{C}_i  $.

By performing  MUSIC algorithms to $ \mathbf{C}_{i,h } $ and $\mathbf{C}_{i,v }$, respectively, we can estimate the AoAs for the LoS paths for each subarray, i.e., $  [  \hat{ \mathbf{k}}_{k,i,0} ^{(c)} ]_2   $ and $ [  \hat{ \mathbf{k}}_{k,i,0} ^{(c)} ]_3  $, $\forall i$. Then, we obtain LoS wave direction  $ \hat{ \mathbf{k}}_{k,i,0} ^{(c)} $ by
 \begin{align}
 	 \left[  \hat{ \mathbf{k}}_{k,i,0} ^{(c)} \right]_1= \sqrt{ 1- \left(   \left[  \hat{ \mathbf{k}}_{k,i,0} ^{(c)} \right]_2   \right)^2 - \left(   \left[  \hat{ \mathbf{k}}_{k,i,0} ^{(c)} \right]_3     \right)^2    }.
 \end{align}

Using the subarray central location and wave vector, we can construct the following linear equations
\begin{align}
	\mathbf{p}_{k,0}^{(c)}  =  \mathbf{p}_{\mathrm{bs},i}  +   r_{k,i,0}^{(c)}  \hat{ \mathbf{k}}_{k,i,0} ^{(c)} , \forall i.
\end{align}
  Then, the sum of the squared distances between  unknown point $ \mathbf{p}_{k,0}^{(c)}  $ and each of the $I$ estimated ray radiated from $ \mathbf{p}_{\mathrm{bs},i} $ towards $  \hat{ \mathbf{k}}_{k,i,0} ^{(c)}  $ is given  by 
\begin{align}
f(  \mathbf{p}_{k,0}^{(c)}    ) \triangleq \! \sum\nolimits_{i=1}^I   \!   \! \left(   \mathbf{p}_{\mathrm{bs},i}     \! -   \!  \mathbf{p}_{k,0}^{(c)}      \right)^{\mathrm{T}} 
\mathbf{B}_{k,i,0}
\left(    \mathbf{p}_{\mathrm{bs},i}      -   \mathbf{p}_{k,0}^{(c)}        \right),
\end{align}
where $\mathbf{B}_{k,i,0}\triangleq \mathbf{I}_3-  \hat{\mathbf{k}}_{k,i,0}  \hat{ \mathbf{k}}_{k,i,0}^{\mathrm{T}}$. Based on  LS criterion, the estimation of location $ \mathbf{p}_{k,0}^{(c)}  $ that minimizes  $ f(  \mathbf{p}_{k,0}^{(c)}    ) $ can be   obtained as
\begin{align}
\hat{ \mathbf{p}  }_{k,0}^{(c)}=\left(\sum\nolimits_{i=1}^I \mathbf{B}_{k,i,0}\right)^{-1}\left(\sum\nolimits_{i=1}^I \mathbf{B}_{k,i,0}    \mathbf{p}_{\mathrm{bs},i}       \right) .
\end{align}

The complexity of stage 2 is nearly  $\mathcal{O}\left(         I\left(      M_{i,h}^3 + M_{i,v}^3   +  M_i^2  \right)                 \right)$  due to MUSIC algorithms.

\subsection{Third Stage:   MIMO Channel Estimation with Location-Aided Dictionary}

Using the estimated location $ \hat{ \mathbf{p}  }_{k,0}^{(c)} $, this stage will construct a small-size location-aided dictionary and estimate the MIMO channel\footnote{Here we only estimate the LoS MIMO channel since if the LoS path exists, it is usually  15-40 dB stronger than the NLoS paths\cite{shi2024double}.   }. We propose to sample a small 3D region around the estimated position $ \hat{ \mathbf{p}  }_{k,0}^{(c)} =[\hat{x}, \hat{y}, \hat{z}]$ in the following way
\begin{align}
\begin{aligned}
	&  {\mathbb{L}}(\hat{x}, \hat{y}, \hat{z})=\big\{(x, y,z) \mid\\
	& \quad x=\hat{x}-\Delta {x}, \hat{x}-\Delta {x}+ {2 \Delta {x} }/{(S_x-1)}, \ldots, \hat{x}+\Delta {x}; \\
	&  \quad y=\hat{y}-\Delta {y}, \hat{y}-\Delta {y}+ {2 \Delta {y} }/{(S_y-1)}, \ldots, \hat{y}+\Delta {y}; \\
	&  \quad z=\hat{z}-\Delta {z}, \hat{z}-\Delta {z}+ {2 \Delta {z} }/{(S_z-1)}, \ldots, \hat{z}+\Delta {z} \big\},
\end{aligned}
\end{align}
where $\Delta {x}, \Delta {y}$, and $\Delta {z}$ represent half the extent of the sampling region along the $x$-, $y$-, and $z$-axes, respectively, with  $S_x$, $S_y$, and $S_z$ denoting the number of sampling points along each axis.  Denote $S=S_xS_yS_z$ as the total number of points sampled in the 3D space. Then, the location-aided dictionary matrix $\mathbf{A}_{\mathbb{L}} \in \mathbb{C}^{MN \times S}$  is constructed as
\begin{align}
\mathbf{A}_{\mathbb{L}}  =  \!\!\left\{     \mathrm{vec}\left(     {\mathbf{H}}_{k} ^{\rm los}  \!  \left(   \mathbf{p}_k\left(x,y,z\right)  \right)    \right)   :    (x,y,z)\in  {\mathbb{L}}(\hat{x}, \hat{y}, \hat{z})    \right\}\!,
\end{align}
where $  {\mathbf{H}}_{k} ^{\rm los}    \left(   \mathbf{p}_k\left(x,y,z\right)  \right)  $ denotes the LoS MIMO channel constructed by the BS array and the user array with a central position of $ \mathbf{p}_k\left(x,y,z\right) $.
 
Recall (\ref{received_signal}) and (\ref{observation_i}). Based on  subarray combing matrix $ \mathbf{V}_{{\rm RF},i,t} $, $\forall i, t$, used in Stage 1, the overall analog combiner employed on XL-MIMO  is $ \mathbf{V}_{{\rm RF}} = [\mathbf{V}_{{\rm {\rm RF}},1}^{\rm T}, \dots,  \mathbf{V}_{{\rm RF}, T}^{\rm T}]^{\rm T} $ where
$ \mathbf{V}_{{\rm RF},t} =\operatorname{blkdiag}\left\{    \mathbf{V}_{{\rm RF},1,t} , \ldots, \mathbf{V}_{{\rm RF},I,t}           \right\} $. Then, the  received signal at the whole XL-MIMO in  block $1$ can be expressed as 
\begin{align}
\mathbf{y}_{k,1} &=  \sqrt{ p_k/N} \;\mathbf{V}_{{\rm RF}} \mathbf{H}_k  \mathbf{w}_{k,1} +  \widehat{\mathbf{V}}_{{\rm RF}} \mathbf{n}_1\\
& = \sqrt{ p_k/N} \left(    \mathbf{w}_{k,1}^{\rm T } \otimes    \mathbf{V}_{{\rm RF}}      \right) \mathrm{vec}\left(    \mathbf{H}_k      \right)   +   \widehat{\mathbf{V}}_{{\rm RF}} \mathbf{n}_1, \\
& \approx \sqrt{ p_k/N} \bar{\mathbf{A}}_{\mathbb{L}} \mathbf{x}_{\mathbb{L}}  +  \widehat{ \mathbf{V}}_{{\rm RF}} \mathbf{n}_1,
\end{align}
where $\bar{\mathbf{A}}_{\mathbb{L}}   =     \left(    \mathbf{w}_{k,1}^{\rm T } \otimes    \mathbf{V}_{{\rm RF}}      \right)  \mathbf{A}_{\mathbb{L}}$
and we can estimate the MIMO channel as  $ {\mathbf{A}}_{\mathbb{L}}  \hat{\mathbf{x}}_{\mathbb{L}} $ after tackling the following CS problem 
\begin{align}
	\begin{aligned}
		\hat{\mathbf{x}}_{\mathbb{L}}  = 	  \arg\min_{ \mathbf{x}_{\mathbb{L}}   }  \left\|    \mathbf{x}_{\mathbb{L}}     \right\|_0
		\text { s.t. }   {\mathbf{y}}_{{k,1}}  = \sqrt{ p_k/N} 
	 \bar{\mathbf{A}}_{\mathbb{L}}        \mathbf{x}_{\mathbb{L}}  .
	\end{aligned}
\end{align}
We want to keep the size of $ \mathbf{A}_{\mathbb{L}} $ small. In this case, OMP might not perform well  without knowing the sparsity feature and under wide grid spacing. To estimate $ \mathbf{x}_{\mathbb{L}}  $ effectively, we propose to employ the powerful tools from sparse Bayesian learning (SBL), which is robust even when the variable is not sparse and does not rely on the information of sparsity. Specifically, SBL can be conducted based on an expectation maximization (EM) algorithm. By treating $    \mathbf{x}_{\mathbb{L}} $ as random vector variables, we assume it follows a Gaussian prior $p\left(\widetilde{\mathbf{x}}_{\mathbb{L}} ; \boldsymbol{\Gamma}_{\mathbb{L}}\right)$ with
\begin{align}
	p\left(\widetilde{\mathbf{x}}_{\mathbb{L}} ; \boldsymbol{\Gamma}_{\mathbb{L}}\right)=\prod_{s=1}^S \frac{1}{\pi \gamma_{s}} \exp\left\{-{\left|\tilde{x}_{ s}\right|^2}/{\gamma_{s} }     \right\},
\end{align}
where $\widetilde{x}_{s} \triangleq\left[\widetilde{\mathbf{x}}_{\mathbb{L}}\right]_s$,$  {\gamma}_{ s}$ denotes the prior variance of $\widetilde{x}_{ s}$, $\boldsymbol{\gamma} \triangleq\left[{\gamma}_{1}, \ldots,{\gamma}_{ S}\right]^{{\rm T }}$, and $\boldsymbol{\Gamma}_{\mathbb{L}}=\operatorname{diag}\left(\boldsymbol{\gamma} \right)$. The  solution of SBL is to find the prior $ p\left(\widetilde{\mathbf{x}}_{\mathbb{L}} ; \boldsymbol{\Gamma}_{\mathbb{L}}  \right) $ that maximize the Bayesian evidence $p\left({\mathbf{y}}_{k,1} ; \boldsymbol{\Gamma}_{{\mathbb{L}}}  \right)$, which leads to the  estimation of $ {\gamma}_{s} $ and $\widetilde{{x}}_{s} $. It can be conducted in an iterative way, with the following EM process:
\begin{align}
	\text{E-step}:  
	\mu_{\widetilde{\mathbf{x}}_{\mathbb{L}} }^{(\ell)} & =\boldsymbol{\Sigma}_{\widetilde{\mathbf{x}}_{\mathbb{L}}   }^{(\ell)} \bar{\mathbf{A}}_{\mathbb{L}}^{ {H}} \bar{\mathbf{y}}_{k,1}  , \\
	\boldsymbol{\Sigma}_{\widetilde{\mathbf{x}}_{\mathbb{L}}}^{(\ell)} & =\left( \bar {\mathbf{A}}_{\mathbb{L}}^{ {H}} \bar{\mathbf{A}}_{\mathbb{L}} +\left(\hat{\boldsymbol{\Gamma}}_{\mathbb{L}}^{(\ell)}\right)^{-1}\right)^{-1} .\\
	\text{M-step}: \widehat{\gamma}_{ s}^{(\ell+1)} &  =\left|\left[\mu_{\widetilde{\mathbf{x}}_{\mathbb{L}}}^{(\ell)}\right]_s\right|^2+\left[\Sigma_{\widetilde{\mathbf{x}}_{\mathbb{L}}}^{(\ell)}\right]_{s, s}, \forall s .
\end{align}
Some intermediate processes are omitted due to the limit space. The complexity of SBL is mainly $\mathcal{O}\left( S^3 \right)$ due to the matrix inverse, which can be light under a small dictionary matrix.

\subsection{Sensing Matrix Design}
To improve the estimation quality of OMP and SBL, we will design the sensing matrix, with the following objective: (1) to make the columns of both  $ \mathbf{V}_{{\rm RF}} $ in (\ref{received_signal}) and $ \mathbf{V}_{{\rm RF},i} $  in (\ref{y_k_i})  mutually orthogonal; and (2) to make the effective noise matrix in (\ref{received_signal})  white and with an un-amplified covariance $\sigma^2$.  

We assume $ T \ge M_i / M_{\mathrm{rf},i} = M_s $ to help sub-connected analog combiner work well.  To begin with, our objective is to design $ \mathbf{V}_{{\rm RF},i} $ to fulfill the condition of
\begin{align}\label{conditionVVI}
	\mathbf{V}_{{\rm RF},i}^H \mathbf{V}_{{\rm RF},i} =   \mathbf{I}_{ M_{ i} }.
\end{align}

Let $\boldsymbol{\mathcal{T}}  \in \mathbb{C}^{ TM_{\mathrm{rf},i} \times TM_{\mathrm{rf},i}  }$ be a permutation matrix that changes the position of the rows of $ \mathbf{V}_{{\rm RF},i}  $ in the following manner: the first row of every $ \mathbf{V}_{{\rm RF},i, t} $, $1\le t \le T$, goes at the top; then, the second row of every $ \mathbf{V}_{{\rm RF},i, t} $  follows and so on. As a result, we have
\begin{align}
	\boldsymbol{\mathcal{T}}  \mathbf{V}_{{\rm RF},i}      =  \mathrm{blkdiag} \left\{    \boldsymbol{F}_1 , \ldots,     \boldsymbol{F}_{M_{\mathrm{rf},i}}        \right\},
\end{align}
where $ \boldsymbol{F}_m  \in \mathbb{C}^{ T  \times  M_s }   $.  Since $ \boldsymbol{\mathcal{T}}^H \boldsymbol{\mathcal{T}} = \mathbf{I} $, condition (\ref{conditionVVI}) holds if $\left( \boldsymbol{\mathcal{T}}\mathbf{V}_{{\rm RF},i} \right)^H\boldsymbol{\mathcal{T}}\mathbf{V}_{{\rm RF},i} =  \mathbf{I}_{ M_{ i} } $, which requires $   \boldsymbol{F}_m^H  \boldsymbol{F}_m =  \mathbf{I}_{M_{ s} }$, $\forall m=[1,M_{\mathrm{rf},i}]$. These conditions can be met based on the DFT matrix. Specifically, generate a  $ T  {M_{\mathrm{rf},i} }\times T {M_{\mathrm{rf},i}} $  DFT matrix $ \boldsymbol{\mathcal{F}}_{T {M_{\mathrm{rf},i}} } $ with constant element modulus $\frac{1}{\sqrt{T}}$. Then, $T$ rows of $\boldsymbol{F}_m $ are set as the  $m, M_{\mathrm{rf},i}+m, 2M_{\mathrm{rf},i}+m, \cdots,  (T-1)M_{\mathrm{rf},i}+m$ rows of  $\left[\boldsymbol{\mathcal{F}}_{T {M_{\mathrm{rf},i}} }\right]_{(:, 1:M_s)}$, respectively, leading to $   \boldsymbol{F}_m^H  \boldsymbol{F}_m =  \mathbf{I}_{M_{ s} }$ and (\ref{conditionVVI}), $\forall m$.
Then, we can also validate that  $	\mathbf{V}_{{\rm RF}}^H \mathbf{V}_{{\rm RF}} =   \mathbf{I}_M $ by using a permutation matrix, and  $\mathbb{E} \{    \widehat{ \mathbf{V}}_{{\rm RF}} \mathbf{n}_1   (  \widehat{ \mathbf{V}}_{{\rm RF}} \mathbf{n}_1 )^H       \} = \sigma^2   \mathbf{I}$ since $\mathbf{V}_{{\rm RF},t}\mathbf{V}_{{\rm RF},t}^H = \mathbf{I}  $, $\forall t$, which achieves our goal.

\section{Numerical Results}
In this section,  numerical results of the proposed localization and MIMO  channel estimation algorithms are presented and compared   with the benchmarks. The center of the BS array is $(0,0,0)$ and the center of the user array is randomly distributed in a rectangular area of $x\in [5,15]$, $y\in[-5,5]$, and $z=-1$. The number of antennas at BS is $M_h=16$ and $M_v=48$ with $M_{{\rm RF}}=256$, $M_{{s}}=6$, forming an aperture of $0.5$ m $\times 1.5$ m. The operation frequency is $6.8$ GHz. The number of antennas per user is $N=4$. We divide the BS array into $I=2\times 4$ tiles.  The parameters about location dictionary are $\Delta x=\Delta y = 0.2$, $\Delta z = 0.02$, $S_x=S_y=11$, and $S_z=3$. The number of combining beams is $T=6$. Metrics of root mean square error (RMSE) and normalized mean square error (NMSE) are used to evaluate the quality of localization  and channel estimation, respectively, with $\operatorname{RMSE}=\sqrt{\mathbb{E} \{\|\hat{\mathbf{p}}_{k,0}^{(c)}-\mathbf{p}_{k,0}^{(c)}\|^2 \}}$ and $\operatorname { NMSE }=\mathbb{E} \{ {\|\hat{\mathbf{H}}_k - \mathbf{H}_k\|_F^2}/{\|\mathbf{H}_k\|_F^2} \}$. 

We consider the following baselines in the simulations:
\begin{itemize}
	\item Antenna-wise SIMO channel estimation using far-field DFT-based dictionary matrix; 
	\item Antenna-wise SIMO channel estimation using spherical-domain dictionary matrix\cite{wu2022multiple}; 
	\item Antenna-wise SIMO channel estimation  applying the far-field DFT dictionary for each subarray;
	\item MIMO channel estimation using location-based eigen-dictionary matrix\cite{liu2025sensing}.
\end{itemize}
Note that for fair comparison, the same overall transmitting power  is employed for all baselines under the different values of $B$. For our methods, there is $B=1$.

\begin{figure}
	\setlength{\abovecaptionskip}{-5pt}
	\setlength{\belowcaptionskip}{-10pt}
	\centering
	\includegraphics[width= 0.4\textwidth]{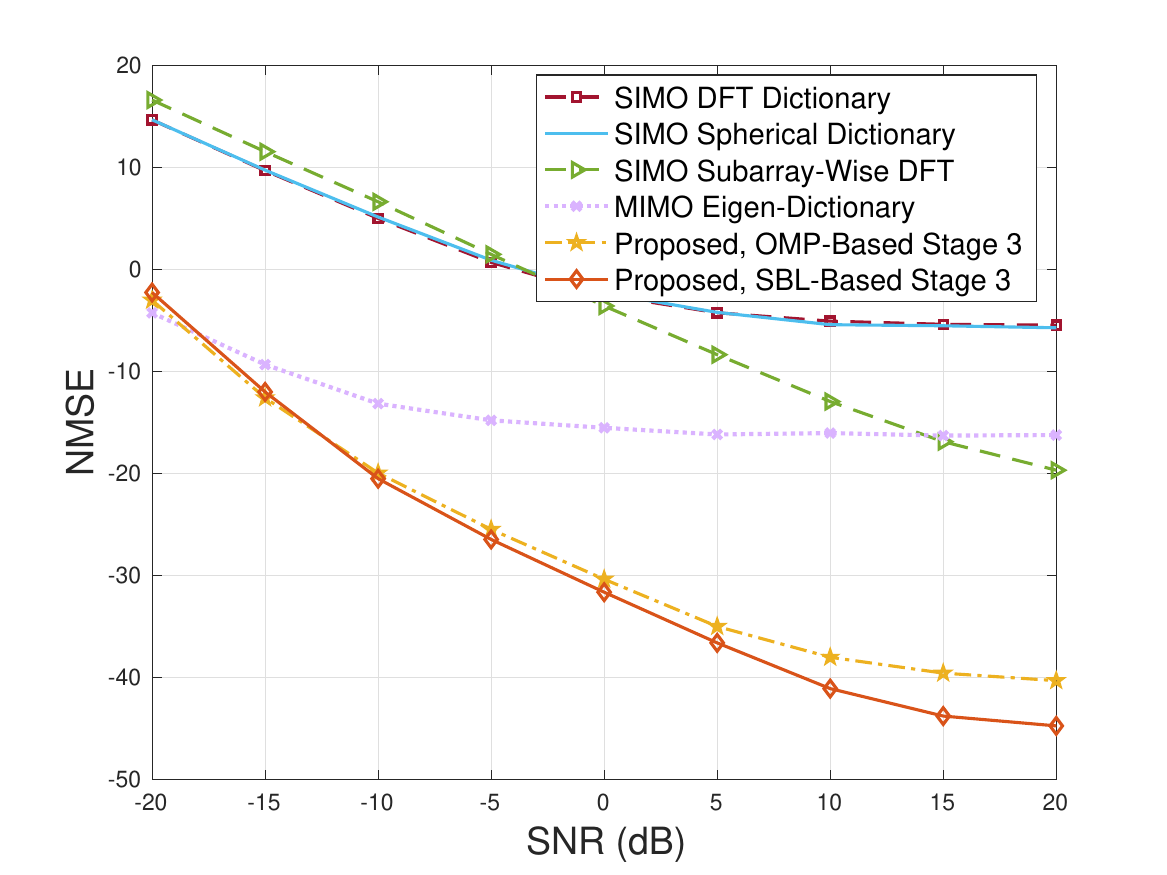}
	\caption{Channel estimation performance.}
	\label{figure2}
	\vspace{-15pt}
\end{figure}

\begin{figure}
	\setlength{\abovecaptionskip}{-5pt}
	\setlength{\belowcaptionskip}{-10pt}
	\centering
	\includegraphics[width= 0.4\textwidth]{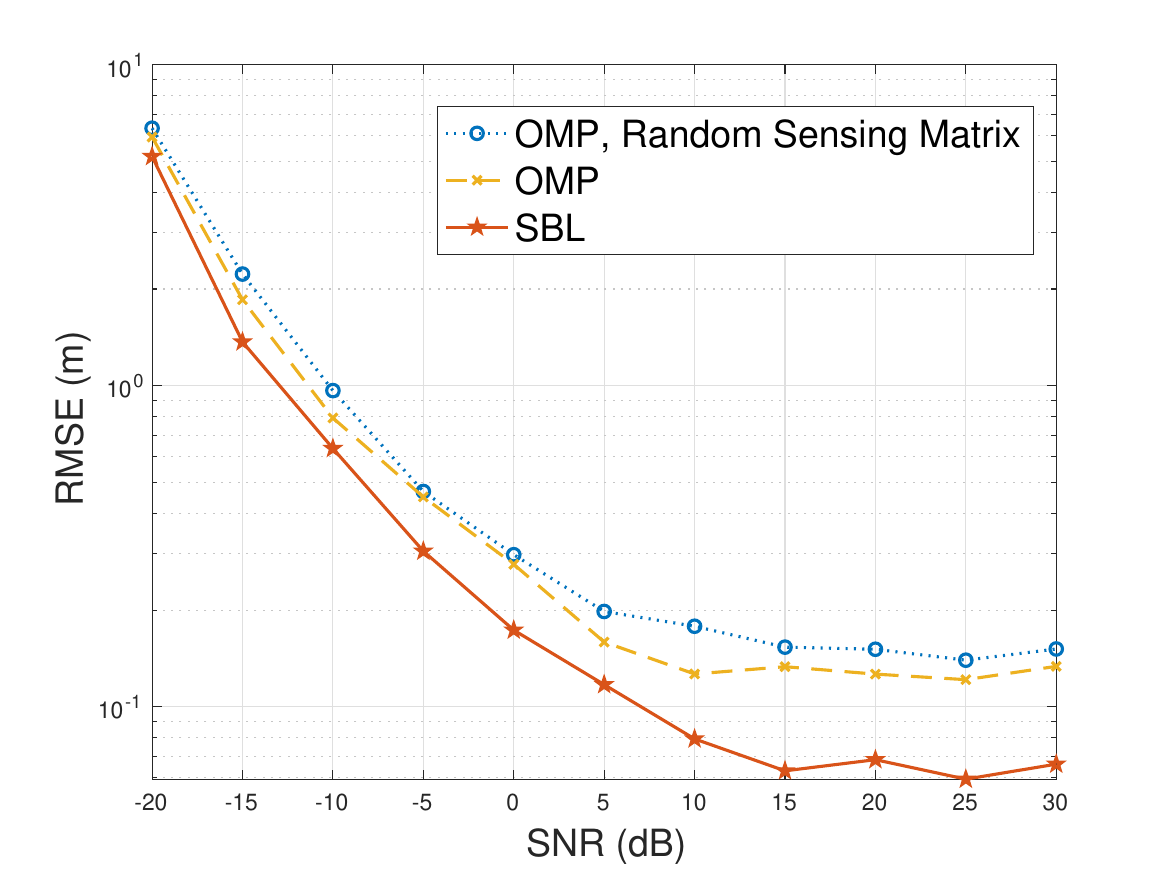}
	\caption{Localization performance.}
	\label{figure3}
	\vspace{-15pt}
\end{figure}

Fig. \ref{figure2} illustrates the MIMO channel estimation performance under different schemes. It can be seen that the proposed method yields a significant advantage compared with all the other benchmarks, which demonstrates the effectiveness of the proposed location-aided dictionary design and the benefits of exploiting the location information in near-field channel estimation. It can be seen that the antenna-wise channel estimation methods using spherical-domain and DFT-based dictionaries can not work well in the considered scenarios since they require a huge dictionary which is hard for CS algorithm to perform effectively. Meanwhile, it can be seen that even using the location information, the method relying on eigen-based dictionary saturates quickly, since it does not fully leverage the possible region around the estimated location. Furthermore, if we only conduct the first stage, i.e., estimating the subarray SIMO channel one by one, the performance is limited due to the inevitable mismatch between near-field channels and the approximated far-field channels. Additionally, by comparing the proposed SBL-based solution to the counterpart where the CS problem in the third stage is solved by OMP, the superiority and robustness of SBL with a small-size dictionary are validated.

Next, in Fig. \ref{figure3}, we investigate the localization performance of the proposed algorithms during the second stage. We plot the results when the subarray channels are estimated by OMP and SBL  in the first stage, respectively. It can be seen that SBL-based subarray channel estimation can lead to centimeter-level localization performance due to the better AoA quality. By contrast, localization using AoAs estimated by the low-complexity OMP algorithm has a coarse precision. Nevertheless, these two localization results actually result in nearly the same MIMO channel estimation quality after stage 3, which demonstrates the robustness of the proposed location-aided dictionary and the feasibility of using OMP in stage 1. Besides, it can be observed that with a random sensing matrix, the localization accuracy will degrade, which showcases the effectiveness of the proposed design schemes about analog combiners $\mathbf{V}_{\mathrm{{\rm RF}}}$.

\section{Conclusion}
This paper proposed a three-stage algorithm to jointly localize the 3D user position and estimate the MIMO channel between multi-antenna user and planar BS array with sub-connected hybrid structures. The subarray channels were estimated in the first stage using OMP, which enabled the user location in the second stage with MUSIC and LS algorithms. In the third stage, a location-aided dictionary matrix was proposed to estimate the MIMO channel  effectively with SBL. 

\bibliographystyle{IEEEtran}
\bibliography{myref.bib}

\end{document}